\documentclass[useAMS,usenatbib,referee]{biom}
\def\bSig\mathbf{\Sigma}

\newcommand{\bX}{\mathbf{X}}

\usepackage{graphicx}
\usepackage{color}
\usepackage{amsmath,amsfonts}
\usepackage[figuresright]{rotating}
\usepackage{threeparttable}
\usepackage{multirow, comment}

\title{Disease Mapping with Generative Models}

\author{Feifei Wang$^{1,*}$\email{wangff@pku.edu.cn},
Jian Wang$^{1,**}$\email{jianw@pku.edu.cn},
Alan E. Gelfand$^{2,***}$\email{alan@stat.duke.edu},
and Fan Li$^{2,****}$\email{fl35@stat.duke.edu} \\
$^{1}$Guanghua School of Management, Peking University, Beijing 100871, P. R. China \\
$^{2}$Department of Statistical Science, Duke University, Durham, NC 27708, U.S.A.}

\begin{document}

\volume{}
\pubyear{}
\artmonth{}

\doi{}

\label{firstpage}

\begin{abstract}
Disease mapping focuses on learning about areal units presenting high relative risk. Disease mapping models for disease counts specify Poisson regressions in relative risks compared with the expected counts. These models typically incorporate spatial random effects to accomplish spatial smoothing. Fitting of these models customarily computes expected disease counts via internal standardization. This places the data on both sides of the model, i.e., the counts are on the left side but they are also used to obtain the expected counts on the right side. As a result, these internally standardized models are incoherent and not generative; probabilistically, they could not produce the observed data. Here, we argue for adopting the direct generative model for disease counts. We model disease incidence instead of relative risks, using a generalized logistic regression. We extract relative risks post model fitting. We also extend the generative model to dynamic settings. We compare the generative models with internally standardized models through simulated datasets and a well-examined lung cancer morbidity data in Ohio. Each model is a spatial smoother and they smooth the data similarly with regard to relative risks. However, the generative models tend to provide tighter credible intervals. Since the generative specification is no more difficult to fit, is coherent, and is at least as good inferentially, we suggest it should be the model of choice for spatial disease mapping.
\end{abstract}


\maketitle

\section{Introduction}
\label{s:introduction}

The mapping of disease incidence and prevalence has a long history in public health and epidemiology (see, e.g., \cite{lawson2000disease,best2005comparison,koch2005cartographies,leyland2005empirical}). A primary objective in disease mapping analysis is to examine the spatial patterns of disease, which can help identify high-risk areas and better allocate health care resources.

Customary disease mapping models assume the disease counts are distributed as Poisson random variables.  For any count, the mean is specified as the product of the expected disease count and relative risk. The expected disease counts are taken as fixed and known quantities, and the relative risks are modeled using generalized linear regressions. Typically, spatial random effects are incorporated to accomplish spatial smoothing of the relative risks.  To compute the expected disease counts, a common practice is \textit{internal} standardization, calculating the expected disease counts as functions of the observed numbers of cases. However, this implies that the data appear on both sides of the model, resulting in a probabilistically incoherent specification. Such models are not generative; probabilistically, they could not produce the data we observe.

The contribution of this paper is to argue on behalf of adopting a coherent generative model specified through disease incidence rather than relative risk, avoiding internal standardization.  Inference regarding relative risks arises as a post model fitting activity.  We introduce spatial smoothing using the usual conditionally autoregressive (CAR) model.  We also extend the generative model to dynamic settings. We compare the generative models with internally standardized models through simulated datasets and also a real dataset.

We find that, by virtue of the flexibility enabled for the random effects under the CAR model,  the internally standardized model and the generative model yield little difference in the point estimation of relative risks. In different words, each model is a spatial smoother and the models smooth the data similarly.  However, the generative model tends to provide tighter credible intervals. We illuminate these results through simulation studies along with an application to a widely investigated dataset on lung cancer mortality in Ohio (\citealt{waller1997hierarchical,xia1997hierarchical,knorr1998modelling,knorr2000bayesian,kottas2008modeling}).  We also show that model fitting for the generative model is no more difficult than for the usual internally standardized model.  Therefore, because it is coherent and is at least as good inferentially, we suggest that it should be the model of choice for spatial disease mapping.

There is a large literature on disease mapping over the past three decades. It essentially dates to \citet{clayton1987empirical}, who built Bayesian Poisson regressions with random intercepts to capture spatial association across relative risks. \citet{besag1991bayesian} split relative risks into different components and present a fully Bayesian framework, which is easily extended to handle more complex settings. Examples include dynamic models considering spatiotemporal effects \citep{waller1997hierarchical,xia1997hierarchical,kottas2008modeling} and multivariate models focusing on regional counts of multiple diseases \citep{knorr2001shared,carlin2003hierarchical,gelfand2003proper,held2005towards}. Again, within this broad literature, the basic formulation models relative risks, using internal standardization to compute the expected disease counts. Therefore, these extensions suffer from the incoherency problem; all can be revised using a generative specification. In particular, we present an illustrative dynamic extension.

Again, in modeling either relative risks or disease incidence, we introduce spatial random effects for smoothing. The most commonly used prior for spatial random effects is the conditional autoregressive (CAR) specification {(\citealt{clayton1987empirical,besag1991bayesian,waller1997hierarchical,knorr1998modelling,banerjee2014hierarchical}). By defining ``neighbors" for each region, the spatial random effects borrow strength locally and thus smooth the local rates toward their neighboring values.  Richer CAR-type models are available (see, e.g., \cite{halloran2000statistical,leroux2000estimation,macnab2000parametric,white2009stochastic}) and can be incorporated in our generative specification.  Here we confine ourselves to the basic CAR model.

The structure of the paper is as follows. Section \ref{s:models} reviews the internally standardized model and introduces the coherent generative model. Section \ref{s:simulation} presents model comparison on simulated data. Section \ref{s:application} presents real data analysis, on both cross-sectional settings and dynamic settings. Section \ref{s:summary} concludes with brief discussion.

\section{Modeling details}
\label{s:models}
\subsection{The Internally Standardized Model}
\label{s:ISmodel}
Suppose for a set of regions $i=1,...,I$, partitioning a study domain, we observe disease counts $Y_i$ as well as a set of region-specific covariates $\bX_i$. Let $p_i$ be the true incidence for region $i$ and $\bar{p}$ be the overall disease rate across the entire study domain. The goal of disease mapping is to infer about the relative risk of the disease, $r_i=p_i/\bar{p}$, for each region. Usually we assume the number of units at risk in each region, $n_i$, is fixed and known and, therefore, $\bar{p} = \frac{\sum_{i} n_{i} p_{i}}{\sum_{i} n_{i}}$. For rare diseases, it is reasonable to use the Poisson approximation to the binomial distribution. The standard model in the literature becomes \citep{clayton1987empirical,besag1991bayesian},
\begin{eqnarray}
Y_i \mid r_i  & \substack{ind\\\sim} & Po(E_i r_i), \nonumber \\
\log(r_i)&=& \bX_{i}^{\prime}\beta + \phi_i, \label{eq:IS_model}
\end{eqnarray}
where the $\phi_i$'s are region-specific random effects, and $E_i$ is the expected number of cases under a null model such as constant risk for all units. The inference goals are to learn about the $\beta$'s but also to implement \emph{smoothing} of the $r_i$ through the $\phi_i$.  While smoothing is desired, there is no notion of a ``best'' smoothing, making model comparison difficult in this regard.

For many datasets, the $E_i$ are computed via \emph{internal standardization} (IS) as
\begin{equation}
E_i=n_i\hat{\bar{p}}=n_i\frac{\sum_{i}Y_i}{\sum_{i}n_i}, \label{IS}
\end{equation}
and are treated as fixed. We refer to Model \eqref{eq:IS_model} as the IS model and, in the sequel, we denote $r_{i}$ in (1) as $r_{i}^{(IS)}$. To incorporate spatial correlation, hence spatial smoothing, across the regions, it is usually assumed that the $\phi_i$'s follow a conditional autoregressive (CAR) distribution \citep{waller1997hierarchical,banerjee2014hierarchical}:
\begin{equation}\label{eq:CAR}
p(\phi_1,\phi_2,...,\phi_{I} \mid \tau)
\propto \tau^{\frac{I}{2}} \exp\{ -\frac{\tau}{2} \sum^{I}_{i=1} \sum_{j<i} w_{ij}(\phi_{i}-\phi_{j})^{2}\},
\end{equation}
where $w_{ij}=1$ if region $j$ is adjacent to region $i$ and $w_{ij}=0$ otherwise. Centering is implemented to identify the $\phi_{i}$'s so  $\sum_{i} \phi_{i} = 0$.

Since the model in (1) is hierarchical, it is often fitted within a Bayesian framework.  Vague Normal  and gamma priors are usually assigned to the $\beta$'s and $\tau$, respectively.  Below we take $\pi(\beta) \propto 1$ and $\tau \sim \text{Gamma(1,1)}$.  The model is usually fit using Markov chain Monte Carlo (MCMC) \citep{besag1991bayesian}.

A concern regarding the IS model is that it is not generative because one cannot obtain the $E_i$'s before $Y_i$'s are realized. In different words, internal standardization implicitly imposes outcome information on both sides of the regression specification, rendering the model probabilistically incoherent. Also, while sensible smoothing may arise from an IS model, failure to acknowledge the uncertainty in the computed $E_i$'s can lead to unsatisfactory assessment of uncertainty in inference, as we show below.

A more satisfying approach to compute $E_i$ is through external standardization, that is, to use an outside data source such as statewide or nationwide data for incidence. However, such data may not exist or, if it does, when the resulting $E_i$'s arise from a different population, collected over a different domain, at a different time, employing them with the $r_{i}^{(IS)}$'s, for the study population in the IS model can still be problematic. An overview of arguments for and against standardization can be found in \citet[][Chapter2]{waller2004applied}.

\subsection{A Coherent Generative Model}
\label{s:CGmodel}
We propose a generative Poisson model for disease mapping with a specification for $p_i$ rather than $r_i$:
\begin{eqnarray}
Y_i \mid p_i & \substack{ind\\\sim} & Po(n_i p_i), \nonumber \\
\mbox{logit}(p_i) &=& \bX_{i}^{\prime}\beta+\phi_{i}, \label{eq:CG_model}
\end{eqnarray}
where, again, the random effects, the  $\phi_i$'s, follow a CAR distribution. We adopt the same priors for the $\beta$ and $\tau$ as in the IS model. The model in (4) can also be fitted straightforwardly using MCMC. It can be viewed as directly applying a Poisson approximation to a binomial model for each region. Additionally, \eqref{eq:CG_model} avoids internal standardization and is coherent; we refer to it as the CG (coherent generative) model hereafter.

If we define $\tilde{r}_i = n_i p_i/E_i$ with $E_i$ as above, then we can recover smoothed $\tilde{r}_i$'s as a post model fitting exercise.  The $r_{i}$ are a linear transformation of the $p_{i}$ and the posterior samples of the $p_i$ immediately provide posterior samples for the $\tilde{r}_i$.  In fact, the `true'' $r_i$ are such that $r_i = \frac{p_{i}}{\bar{p}}$ where, again, $\bar{p}= \frac{\sum_{i} n_{i}p_{i}}{\sum_{i} n_{i}}$.  So, the smoothed true $r_i$ can also be recovered post model fitting.  Posterior samples of the $p_i$ will provide posterior samples of the $r_i$ but now the transformation is nonlinear.  In the sequel we denote these two choices of relative risks as $\tilde{r}_{i}^{(CG)}$ and $r_{i}^{(CG)}$. Again, $\tilde{r}_{i}^{(CG)}= \frac{p_{i}}{\hat{\bar{p}}}$ while $r_{i}^{(CG)} = \frac{p_{i}}{\bar{p}}$. Below, we compare the inference regarding these relative risks along with the $r_{i}^{(IS)}$.

We offer a simple illumination of the adjustment provided by the spatial random effects. In the CG model, $p_{i}=\frac{\exp(\bX_i^{\prime}\beta+\phi_i)}{1+\exp(\bX_i^{\prime}\beta+\phi_i)}$. Let $p_{i,non}=\frac{\exp(\bX_i^{\prime}\beta)}{1+\exp(\bX_i^{\prime}\beta)}$ and $\eta_i=e^{\phi_i}$, then we can rewrite $p_i$ as $p_i=\frac{p_{i,non}}{p_{i,non}+(1-p_{i,non})/\eta_i}$.
Values of the $\eta_i$'s determine the degree of spatial smoothing of the $p_i$'s: there is no smoothing of $p_i$'s when $\eta_i=1$ (corresponding to $\phi_i=0$); $p_i$'s are smoothed upward when $\eta_i>1$ and downward when $\eta_i<1$. Similarly in the IS model, we can write $r_i=\eta_ir_{i,non}$, where $r_{i,non}=\exp(\bX_{i}'\beta)$; the values of the $\eta_i$'s play a similar role in smoothing the $r_i$'s.

If we summarize the relative risks in terms of posterior means, we find the following.  Let $\eta_{i}^{(IS)} = e^{\phi_{i}^{(IS)}}$.  Then, $E(r_{i}^{(IS)}|\texttt{data}) = E(\eta_{i}^{(IS)}\exp\{\bX_i^{\prime}\beta^{(IS)}\}|\texttt{data})$.  Similarly, let $\eta_{i}^{(CG)} = e^{\phi_{i}^{(CG)}}$. Given that $\exp\{\bX_i^{\prime}\beta^{(CG)}+\phi_{i}^{(CG)}\}$ is usually small, we have
\[E(\tilde{r}_{i}^{(CG)}|\texttt{data}) \approx
\frac{n_i}{E_i}E(\eta_{i}^{(CG)}\exp\{\bX_i^{\prime}\beta^{(CG)}\}|\texttt{data}).\]  The flexibility in scaling of the $\eta$'s suggests that there will be little difference in these two estimated relative risks.

Because of the nonlinear form in the $p_{i}$'s, $E(r_{i}^{(CG)}|\texttt{data})$ is difficult to accurately assess analytically.  However, its behavior can be quite different from $E(\tilde{r}_{i}^{(CG)}|\texttt{data})$.  For example, if the model for $p_{i}$ is $\text{logit}(p_{i}) = \beta_{0}$, i.e., $p_i$ is constant over all regions, then $r_{i}^{(CG)}=1$ for all $i$, regardless of the data.  When a constant $p_i$ model is inappropriate for the data, then, evidently, it is an inappropriate model for learning about relative risk.  However, $E(\tilde{r}_{i}^{(CG)}|\texttt{data}) = E(\frac{e^{\beta_{0}}}{1+ e^{\beta_{0}}}|\texttt{data})/\hat{\bar{p}}$,  which need not be close to 1 but will be close to $E(r_{i}^{(IS)}|\texttt{data})$.  The posterior variances for these relative risks are not accessible analytically but are examined in the simulation and data analysis examples below.

Rates of disease incidence, the $p_i$'s, are often very small and therefore link functions distinguishing small probabilities may be more appropriate than the standard logit link. Therefore, we investigated two alternative links -- the complementary log-log (c-loglog) link, which approaches 0 fairly slowly,
\begin{equation}
p_i\ =\ 1-\exp\{-\exp(\bX_i^{\prime}\beta+\phi_{i})\},
\end{equation}
and the skewed logit link,
\begin{equation}
p_i\ =\ \frac{c_0\{\exp(\bX_i^{\prime}\beta+\phi_{i})\}}{1+ c_{0}\exp(\bX_i^{\prime}\beta+\phi_{i})},
\end{equation}
which, for suitable $c_{0}$, rises faster than the usual logit and thus also may help to distinguish small values better.

\subsection{Dynamic Model Specifications}
The CG model can be readily extended to dynamic settings, where the primary objective is to explain the spatiotemporal patterns of disease and perhaps to make (smoothed) forecasts of future disease risks. However, smoothed prediction implies that \emph{accurate} future prediction is not an objective.  In building such spatiotemporal models, one needs to specify spatial effects, temporal effects and their interactions. A variety of formulations have been proposed, including additive models \citep{knorr1998modelling}, multiplicative models \citep{bernardinelli1995bayesianb, xia1997hierarchical}, independent CAR models for different time periods \citep{waller1997hierarchical}, or the general \textit{spatiotemporal autoregressive} (STAR) models \citep{pace2000method}.  Illustratively, we consider a simple additive model for the spatiotemporal effects without interaction, where the temporal random effects follow an AR(1) model and the spatial random effects follow a CAR model. Letting $Y_{it}$ be the disease counts for region $i$ at time $t$ for $i=1,...,I$ and $t=1,..., T$, the dynamic CG model is,
\begin{eqnarray}\label{eq:CG_dynamic}
Y_{it} &\sim& Po(n_{it} p_{it}), \nonumber \\
\mbox{logit}(p_{it})&=& \bX'\beta + \phi_i + \alpha_t, \nonumber \\
\alpha_t &=& \rho \alpha_{t-1} + \delta_{t}, \nonumber \\
 (\phi_1,\phi_2,...,\phi_{I}) &\sim& \mbox{CAR}(\tau),
\end{eqnarray}
again with $\pi(\beta) \propto 1$, $\tau \sim \mbox{Gamma}(1,1)$, and $\delta_{t} \sim \mbox{N}(0,\omega)$ with $\pi(\rho,\omega) \propto \omega^{-1}$.

For comparison, we can also extend the IS model to a similar dynamic setting:
\begin{eqnarray} \label{eq:IS_dynamic}
Y_{it} &\sim& Po(E_{it} r_{it}) \nonumber \\
\log(r_{it}) &=& \bX'\beta + \phi_i + \alpha_t
\end{eqnarray}
with the same AR(1) model for the temporal effects $\alpha_t$ and CAR model for the spatial effects $\phi_i$ as in the CG model \eqref{eq:CG_dynamic}. Similarly to the above, the expected disease counts $E_{it}$ are calculated via internal standardization at each time as $E_{it}=n_{it}\sum_{i}Y_{it}/\sum_{i}n_{it}$.

\section{Simulation examples}
\label{s:simulation}

\subsection{Simulation Design}

We design our simulation investigation based upon a real application which we take up in Section 4 -- the lung cancer mortality data from the state of Ohio. The data was originally extracted from a public use database (Centers for Disease Control, 1988) for every county in the United States.  It has been widely studied in disease mapping \citep{waller1997hierarchical,xia1997hierarchical,knorr1998modelling,knorr2000bayesian,kottas2008modeling}. In Section 4 we focus on a subset of $I=88$ Ohio counties over a period of $T=21$ years (from 1968 to 1988), yielding a total of 7392 observations.

For our simulation, we use the same geographical map of the 88 Ohio counties and the population data for the year 1988. To somewhat mimic the real spatial scenario, we assign the ``true" incidence rates $p_i$ for these counties as follows: we first set $p_i=0.001$ for all counties, and then for the three most populated counties, Cuyahoga county (containing Cleveland), Franklin county (containing Columbus) and Hamilton county (containing Cincinnati), we assign higher incidence rates by increasing $p_i$ by 0.0015, 0.001 and 0.001 respectively. Finally for all adjacent counties of these three counties, we increase $p_i$ by 0.0005. Figure \ref{f:truep} depicts the resulting map of the true $p_i$'s.

\begin{figure}
    \centerline{\includegraphics[width=4in]{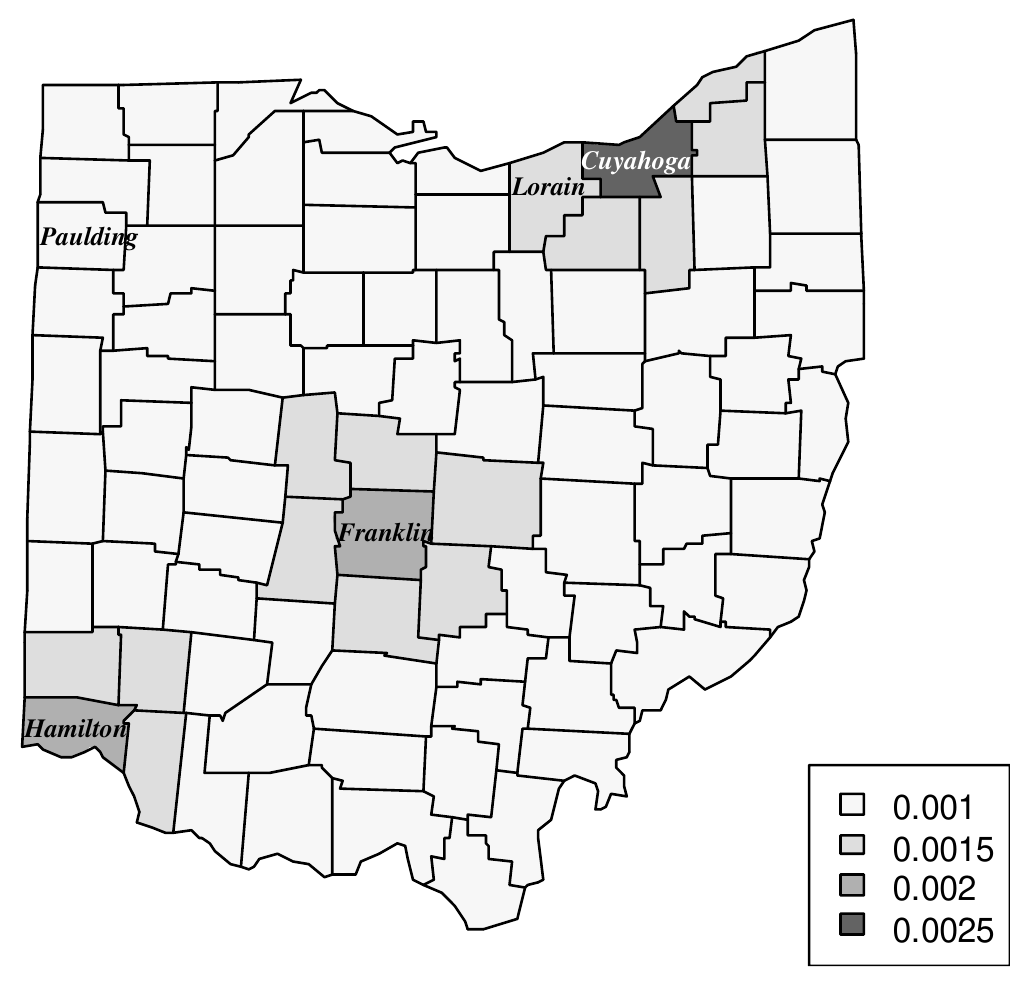}}
    \caption{Map of the true disease incidence $p_i$ in the simulation study.}
    \label{f:truep}
\end{figure}

Using these assigned values of $p_i$'s and the true $n_i$'s in the real Ohio data, we generate $Y_i$ from $Po(n_{i}p_{i})$. We repeated this $B=500$ times to create $500$ simulated datasets.

\subsection{Comparison Criteria}
\label{s:comparison}
For model comparison, we focus on inference regarding the $r_i$'s -- the main objective in the IS model.  We compare the point estimates of the two models using two loss functions motivated by counts: the relative squared error, $\mbox{ratio}(r)=(\hat{r}_i-r_i)^2 /r_i$, and the squared \emph{bias} of the logarithm of the $r_i$'s, $\mbox{bias}(r)=\{\log(\hat{r_i})-\log(r_i)\}^2$. We also compare the empirical coverages and lengths of the nominal 90\% credible intervals to evaluate the uncertainty of the estimates from each model.

Note that since smoothing is the goal of the disease mapping modeling, it is not sensible to assess model performance in terms of recovery of the raw counts or rates; we could achieve this perfectly without smoothing. Rather, we need to think about performance in the sense of \emph{shrinkage} estimation \citep{efron1973stein,efron1977stein}, i.e., in terms of expected loss for estimating the vector of mean relative risks.  That is, we know that, in three or more dimensions, certain shrinkage estimators dominate the maximum likelihood estimators in terms of risk for certain choices of loss functions.  For count data, the foregoing choices are suitable to consider.

As a result, for loss function, $L(\hat{r}_{i}, r_{i})$ where $r_i$ is the true relative risk, we need to study $E(\sum_{i} L(\hat{r}_{i}, r_{i}))$, the overall expected loss, for an estimator $\{\hat{r}_{i}, i=1,2,..I\}$.  We can do this through simulation, i.e., by obtaining a Monte Carlo integration for the expectation, through generation of samples under a specific set $\{p_{i}, i=1,2...I\}$ (which induces a set of $r_i$'s), computing the loss for each sample and averaging over the samples.  In this regard, out of sample prediction in space or time is not an appropriate way to assess the performance of a finite dimensional model whose intent is smoothing.

\subsection{Simulation Results}
We first present results from a randomly chosen single simulated dataset. Figure \ref{f:est_r} displays the spatial surfaces of the three $r_i$'s estimated from the two models incorporating spatial effects. As expected, the spatial maps are nearly identical, with counties around Cuyahoga, Franklin and Hamilton having higher relative risks than the others. An exception is Lorain (see Figure \ref{f:truep}), which is identified as one of the high risk counties ($r_i>1$) by using $r^{(CG)}_{i}$, but not the case by using $r^{(IS)}_{i}$ and $\widetilde{r}^{(CG)}_{i}$. As Lorain has very high risk with true relative risk being 1.023, this suggests that the CG model with estimator $r^{(CG)}_{i}$ may be better in discovering high risk counties.

\begin{figure}
    \centerline{\includegraphics[width=7in]{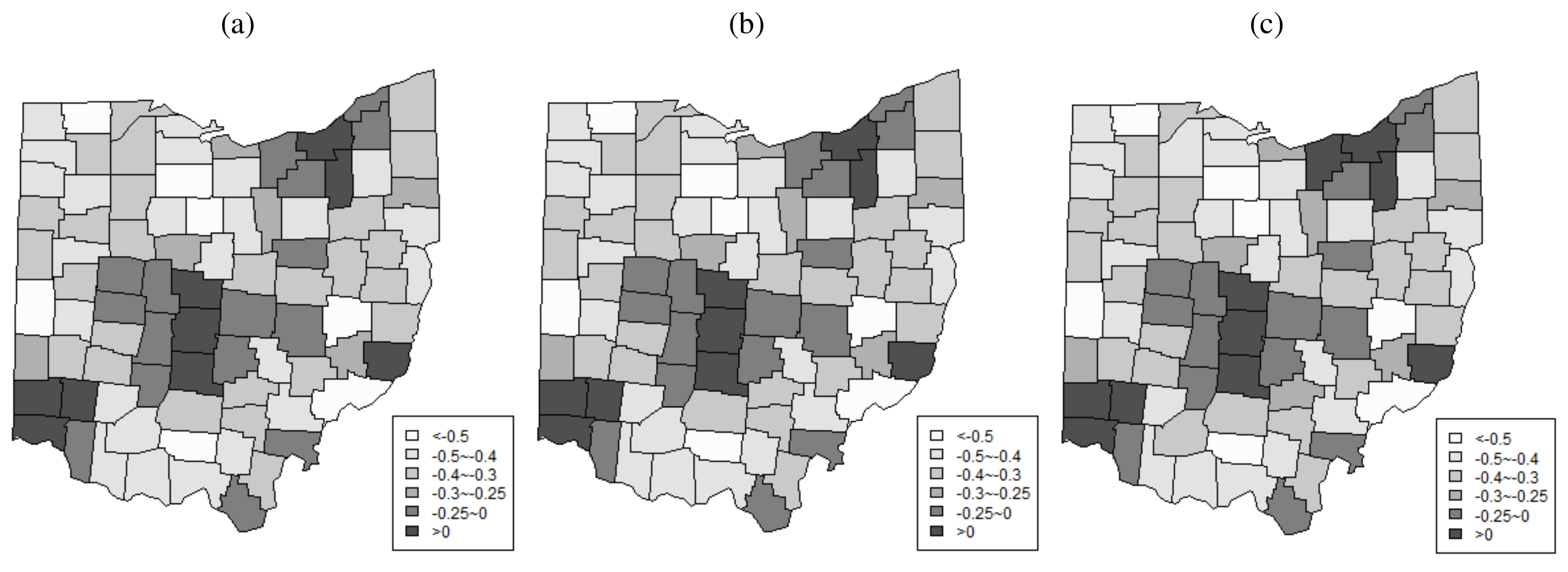}}
    \caption{Maps of three estimators of the logarithm of $r_i$'s. (a, b, c represent $r^{(IS)}_{i}$, $\widetilde{r}^{(CG)}_{i}$ and $r^{(CG)}_{i}$ respectively.)}
    \label{f:est_r}
\end{figure}

To further examine the smoothing performance, we show shrinkage for each of the three estimators in Figure \ref{f:shrink}. In each shrinkage plot, we compare the model estimates incorporating spatial effects with the MLE estimates $Y_{i}/E_{i}$. We find the three shrinkage plots are nearly identical, and all of them have substantially shrunk the model estimates towards the grand mean for the counties with very low or high MLE estimates.

\begin{figure}
    \centerline{\includegraphics[width=6in]{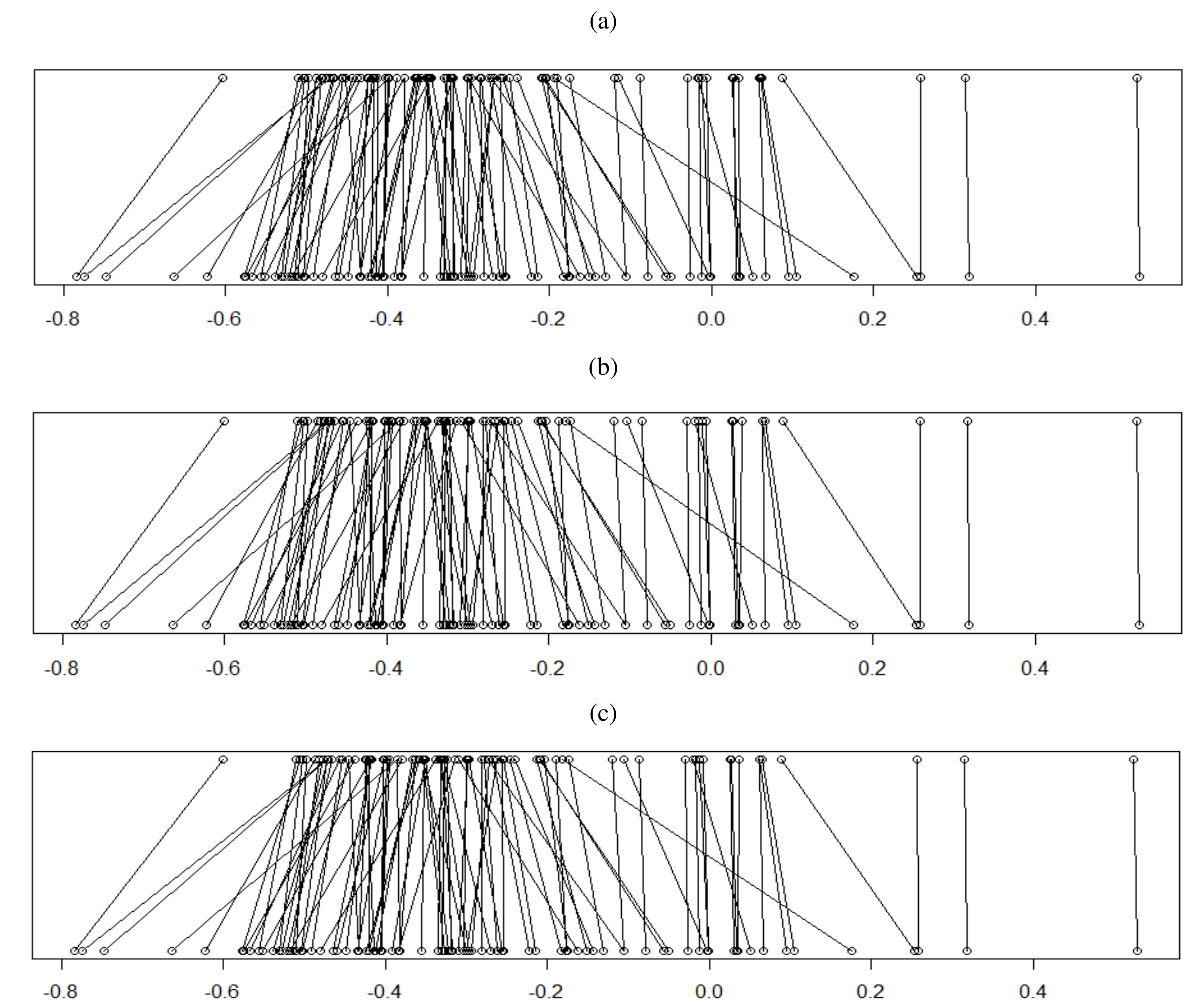}}
    \caption{Shrinkage plots for three estimators of the logarithm of $r_i$'s. (a, b, c represent $r^{(IS)}_{i}$, $\widetilde{r}^{(CG)}_{i}$ and $r^{(CG)}_{i}$ respectively. In each panel, the upper points represent the model estimates, while the lower points represent the MLE estimates.)}
    \label{f:shrink}
\end{figure}

We then compare the three estimators along with the MLE estimates through expected loss, which is calculated using the 500 replicated samples. The expected losses for $r_i^{(IS)}$, $\widetilde{r}_i^{(CG)}$, $r_i^{(CG)}$ and $Y_i/E_i$ are 0.5486, 0.5490, 0.5485, and 1.3268 respectively when using ratio($r$); the corresponding expected losses are 0.7375, 0.7378, 0.7371 and 2.0364 when using bias($r$). Though the three model estimates show no distinguishable differences, they all substantially outperform the MLE estimates.

Next, we compare inference under the three model estimates in terms of uncertainty. For a given model, each of the 500 samples provides a $90\%$ credible interval for the relative risk for each of the $88$ counties.  Each interval yields a length and a binary variable recording whether it contained the true value (1) or not (0).  Hence, we can create a $500 \times 88$ matrix of lengths as well as a $500 \times 88$ matrix of binary outcomes.  We can examine these matrices by rows, by columns, or overall.  We seek to compare them across the results for the three model estimates.
Summary results are present in Table~\ref{t:simures}. Under the actual population sizes, the $n_i$'s, and the usual logit link, the resulting expected coverage probabilities, averaged over the $88$ counties, exceed the nominal probabilities for each of the three estimators.  They are 94.39\%, 94.47\% and 94.14\% and the overall average length are 0.2538, 0.2537 and 0.2517, respectively. To make row-wise comparison for the two $CG$ model estimators with the $IS$ model, we calculate the percentage of time (out of $500$ replications) the $CG$ interval was shorter than the $IS$ model. For $\widetilde{r}_i^{(CG)}$ we find 51\% of their average intervals are shorter than $r_i^{(IS)}$, while for $r_i^{(CG)}$, we find 95.2\% are shorter than $r_i^{(IS)}$.
To make column-wise comparison for the two $CG$ model estimators with the $IS$ model, we calculate the proportion of times (out of $88$) the $CG$ interval was shorter than the $IS$ model. For $\widetilde{r}_i^{(CG)}$ we find 48.86\% of their average intervals are shorter than $r_i^{(IS)}$, while for $r_i^{(CG)}$ we find 88.64\% shorter than $r_i^{(IS)}$. These findings suggest that the CG model with estimator $r_i^{(CG)}$ tends to  achieve tighter credible intervals than the IS model, while $\widetilde{r}_i^{(CG)}$ and $r_i^{(IS)}$ do not show much difference.
\begin{table}
  \caption{Simulation results of three estimators, under different population size and different links.}
  \label{t:simures}
  \begin{center}
  \begin{threeparttable}[b]
  \begin{tabular}{ccccccc}
  \Hline
  \multirow{2}{*}{Population} &\multirow{2}{*}{Link} &\multirow{2}{*}{Estimator}
  &\multicolumn{2}{c|}{Average of $500\times 88$} &\multicolumn{2}{c|}{\% $(\mbox{CI}^{(CG)}< \mbox{CI}^{(IS)})$}\\
  \cline{4-7}
  & & &Coverage &Length &Row-wise &Column-wise\\
  \hline
  \multirow{9}{*}{$n_i$} &\multirow{3}{*}{usual logit}
  &$r_i^{(IS)}$	            &94.39\%	&0.2538	&$-$	&$-$	\\
  & &$\widetilde{r}_i^{(CG)}$	&94.47\%	&0.2537	&51.00\%	&48.86\%	\\
  & &$r_i^{(CG)}$	            &94.14\%	&0.2517	&95.20\%	&88.64\%	\\
  \cline{2-7}
  &\multirow{3}{*}{c-loglog}
  &$r_i^{(IS)}$	            &94.25\%	&0.2536	&$-$	&$-$	\\
  & &$\widetilde{r}_i^{(CG)}$	&94.37\%	&0.2538	&42.75\%	&31.82\%	\\
  & &$r_i^{(CG)}$	            &94.00\%	&0.2517	&93.25\%	&87.50\%	\\
  \cline{2-7}
  &\multirow{3}{*}{skewed logit}
  &$r_i^{(IS)}$	            &94.15\%	&0.2538	&$-$	&$-$	\\
  & &$\widetilde{r}_i^{(CG)}$	&94.27\%	&0.2541	&41.60\%	&34.09\%	\\
  & &$r_i^{(CG)}$	            &94.10\%	&0.2519	&93.20\%	&86.36\%	\\
  \hline
  \multirow{9}{*}{$n_i/10$} &\multirow{3}{*}{usual logit}
  &$r_i^{(IS)}$	            &97.54\%	&0.5568	&$-$	&$-$	\\
  & &$\widetilde{r}_i^{(CG)}$	&97.49\%	&0.5567	&52.00\%	&55.68\%	\\
  & &$r_i^{(CG)}$	            &97.19\%	&0.5512	&90.40\%	&97.73\%	\\
  \cline{2-7}
  &\multirow{3}{*}{c-loglog}
  &$r_i^{(IS)}$	            &97.50\%	&0.5583	&$-$	&$-$	\\
  & &$\widetilde{r}_i^{(CG)}$	&97.50\%	&0.5583	&50.20\%	&52.27\%	\\
  & &$r_i^{(CG)}$	            &97.20\%	&0.5528	&89.00\%	&95.45\%	\\
  \cline{2-7}
  &\multirow{3}{*}{skewed logit}
  &$r_i^{(IS)}$	            &97.34\%	&0.5564	&$-$	&$-$	\\
  & &$\widetilde{r}_i^{(CG)}$	&97.34\%	&0.5566	&49.40\%	&40.91\%	\\
  & &$r_i^{(CG)}$	            &97.04\%	&0.5512	&88.00\%	&94.32\%	\\
  \hline
  \end{tabular}
  \end{threeparttable}
  \end{center}
\end{table}

We then replace the usual logit link with the c-loglog link and with the skewed logit link; the results are also shown in Table~\ref{t:simures}. For the skewed logit link, we set $c_0=0.004$ to make it favor small probabilities. Under the population size $n_i$, the c-loglog link and the skewed logit link yield similar results to those for the usual logit link. The resulting averages of the expected coverage probabilities are all around $94\%$.  With regard to the length of credible intervals, $\widetilde{r}_i^{(CG)}$ and $r_i^{(IS)}$ are comparable, but $r_i^{(CG)}$ tends to result in shorter intervals than $r_i^{(IS)}$ in both row-wise and column-wise comparison.

To consider sensitivity to population sizes, we reduce the actual size of $n_i$ to $n_i/10$ and then redo the entire simulation exercise. As shown in Table~\ref{t:simures}, it seems that now the resulting expected average of the coverage probabilities rises to roughly $97\%$.  Also, regardless of the links, $r_i^{(CG)}$  leads to tighter credible intervals than $r_i^{(IS)}$, with both the row-wise and column-wise proportions  around 90\%. The estimator $\widetilde{r}_i^{(CG)}$ behaves similarly to $r_i^{(IS)}$. This suggests robustness of the performance of the CG model to the population size.

\section{Application to the Ohio cancer mortality data}
\label{s:application}
\subsection{Cross-sectional settings}
We now apply the two models with spatial random effects, yielding three estimators, to analyze the Ohio lung cancer mortality data. For the static analysis we consider relative risks for the first and last years, 1968 and 1988. Figure \ref{f:logr} presents maps of the raw estimates ($Y_i/E_i$) and the three model estimates of the $r_i$'s in the two specific years, respectively. For each year, the spatial estimates of the models are nearly identical and all three reveal the smoothing of the raw rates. Counties with extreme raw relative risks are smoothed by their neighbors. An example is Paulding county (see Figure \ref{f:truep}), which has a high raw rate in year 1968 but is adjacent to several low risk counties; due to the spatial effects, the models  substantially reduce the estimated relative risk for this county.
\begin{figure}
    \centerline{\includegraphics[width=7in]{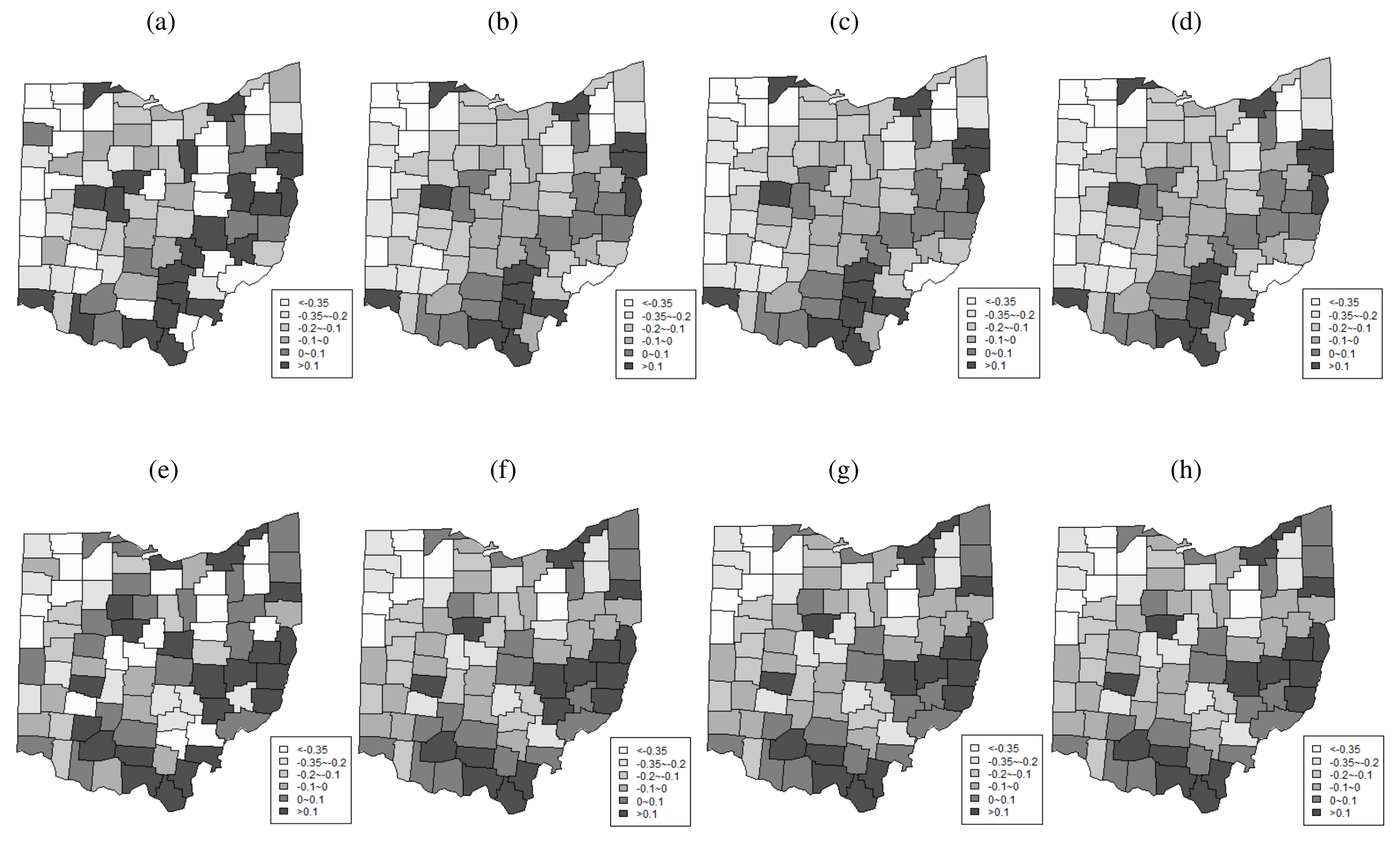}}
    \caption{Maps of different estimators of the $r_i$'s (in logarithm) for Ohio data in year 1968 and 1988. (a, b, c d are raw rates, $r_i^{(IS)}$, $\widetilde{r}_i^{(CG)}$ and $r_i^{(CG)}$ for year 1968; e, f, g, h are corresponding values for year 1988.)}
    \label{f:logr}
\end{figure}

The CG model with estimator $r_i^{(CG)}$ results in tighter credible intervals than the IS model in 62 and 60 out of the 88 counties in year 1968 and 1988, respectively. For $\widetilde{r}_i^{(CG)}$, the number of shorter credible intervals than the IS model is only 50 and 49 out of the 88 counties in year 1968 and 1988, respectively. In particular, visible from Figure \ref{f:lengthr}, the CG model with estimator $r_i^{(CG)}$ tends to produce shorter intervals for several counties located in northeastern and southern Ohio.
\begin{figure}
    \centerline{\includegraphics[width=6in]{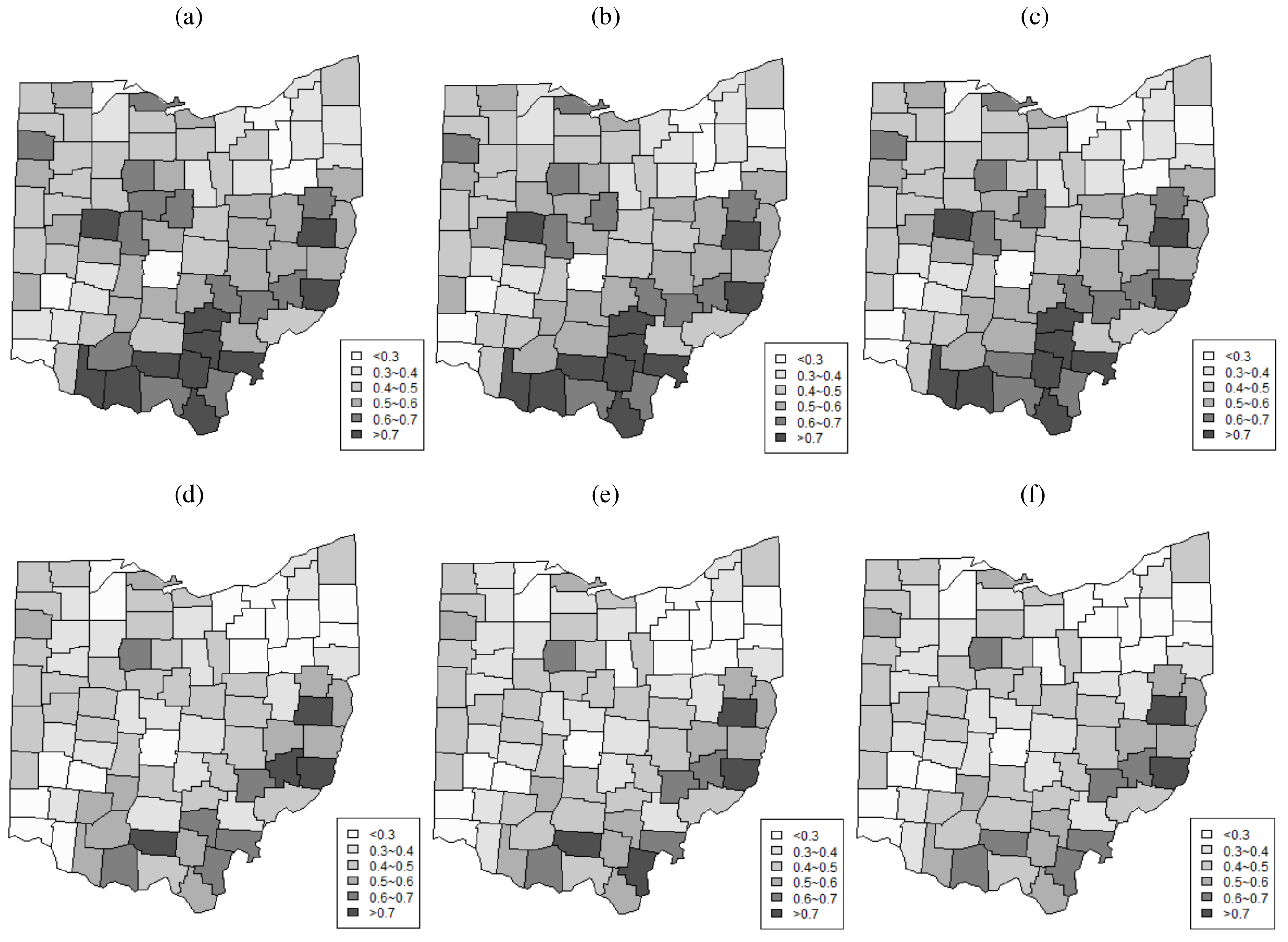}}
    \caption{Maps of the length of 90\% credible intervals of the $r_i$'s for Ohio data in year 1968 and 1988. (a, b, c are $r_i^{(IS)}$, $\widetilde{r}_i^{(CG)}$ and $r_i^{(CG)}$ for year 1968, and d, e, f are those for year 1988.)}
    \label{f:lengthr}
\end{figure}

\subsection{Dynamic settings}
We fit the two dynamic models to the Ohio data consisting of $T=21$ years, one for the full dataset and one for the subset of white males between 55 and 64 years \citep{knorr2000bayesian,kottas2008modeling}. Holding out year 1988, we fit models for the $20$ years from 1968 to 1987.  We assess model performance for those $20$ years and then make predictions for relative risks in year 1988. Perhaps the most important finding is the magnitude of $\hat{\rho}$ in Table~\ref{t:dynamicres}.  Particularly for the full dataset, there is very strong first order temporal correlation. To assess the benefit of the spatio-temporal model, we consider year 1987. We first compare the lengths of 90\% credible intervals of the three estimators, say $r_i^{(IS)}$, $\widetilde{r}_i^{(CG)}$ and $r_i^{(CG)}$, in dynamic models. We find the dynamic CG model with estimator $r_i^{(CG)}$ again tends to provide tighter intervals for the $r_i$'s than the IS model: 57 and 55 out of the 88 counties for the full dataset and the subset, respectively. The estimator $\widetilde{r}_i^{(CG)}$ is comparable to $r_{i}^{(IS)}$; only 45 and 42 out of the 88 counties have shorter intervals for the full dataset and the subset, respectively.

We then compare the results for the dynamic model with those for the static model for year 1987; the results are shown in Table~\ref{t:dynamicres}. For each of the three estimators, the dynamic models yield smaller average length of credible intervals than the static models. Also, when making a pair-wise comparison for each county, all the 88 counties have tighter credible intervals from the dynamic models than from the static models. These findings suggest that gathering $20$ years of data into the dynamic models reduces uncertainty relative to that for the single year static models.

Because the goal of dynamic models is to provide explanation and smoothing in space and time rather than prediction, we provide prediction results solely for information. We use both predictive mean square error (PMSE) and continuous rank probability scores (CRPS) \citep{gneiting2007strictly} to compare the predicted values and predictive distributions with the observed raw risks. We also examine the length of 90\% credible intervals and their coverage for the observed raw rates. As shown in Table~\ref{t:dynamicres}, the three estimators $r_{i}^{(IS)}$, $\widetilde{r}_{i}^{(CG)}$ and $r_{i}^{(CG)}$ result in similar MSE and CRPS values, and the coverage rates are all very low.

\begin{table}
  \caption{Results for dynamic analysis of the Ohio data (D and S present dynamic and static models)}
  \label{t:dynamicres}
  \begin{center}
  \begin{threeparttable}[b]
  \begin{tabular}{ccccccccc}
  \Hline
  \multirow{2}{*}{Dataset} &\multirow{2}{*}{Estimator} &\multirow{2}{*}{$\hat{\rho}$}
  &\multicolumn{3}{c|}{Length of CIs (1987)} &\multicolumn{3}{c|}{Prediction (1988)}\\
  \cline{4-9}
  & & &Avg.(D) &Avg.(S) &\% (D$<$S) &PMLE &CRPS &Coverage \\
  \hline
  \multirow{3}{*}{Full}
  &$r_i^{(IS)}$	            &0.9344	&0.2743	&0.3940	&100\%	&0.2204	&0.2327	&39.77\%	\\
  &$\widetilde{r}_i^{(CG)}$	&0.9341	&0.2749	&0.3972	&100\%	&0.2206	&0.2327	&40.91\%	\\
  &$r_i^{(CG)}$	            &0.9341	&0.2746	&0.3948	&100\%	&0.2206	&0.2327	&40.91\%	\\
  \hline
  \multirow{3}{*}{Subset}
  &$r_i^{(IS)}$	            &0.7569	&0.3818	&0.5936	&100\%	&0.4967	&0.5034	&38.64\%	\\
  &$\widetilde{r}_i^{(CG)}$	&0.7686	&0.3838	&0.6025	&100\%	&0.4966	&0.5032	&38.64\%	\\
  &$r_i^{(CG)}$	            &0.7686	&0.3847	&0.6114	&100\%	&0.4966	&0.5032	&38.64\%	\\
  \hline
  \end{tabular}
  \end{threeparttable}
  \end{center}
\end{table}

We select two counties in the subset to display the spatiotemporal effects. One is Hamilton, a highly populated county located in southwestern part of the state, and the other is Delaware, more suburban and located in central Ohio. Figure \ref{f:trend} presents the estimated relative risks and their 90\% credible intervals obtained from the CG model using estimator $r_i^{(CG)}$ for year 1968 to 1988. For both counties, the dynamic CG model has smoothed the raw rates to suggest an increasing trend in relative risk over time.
\begin{figure}
    \centerline{\includegraphics[width=5.5in]{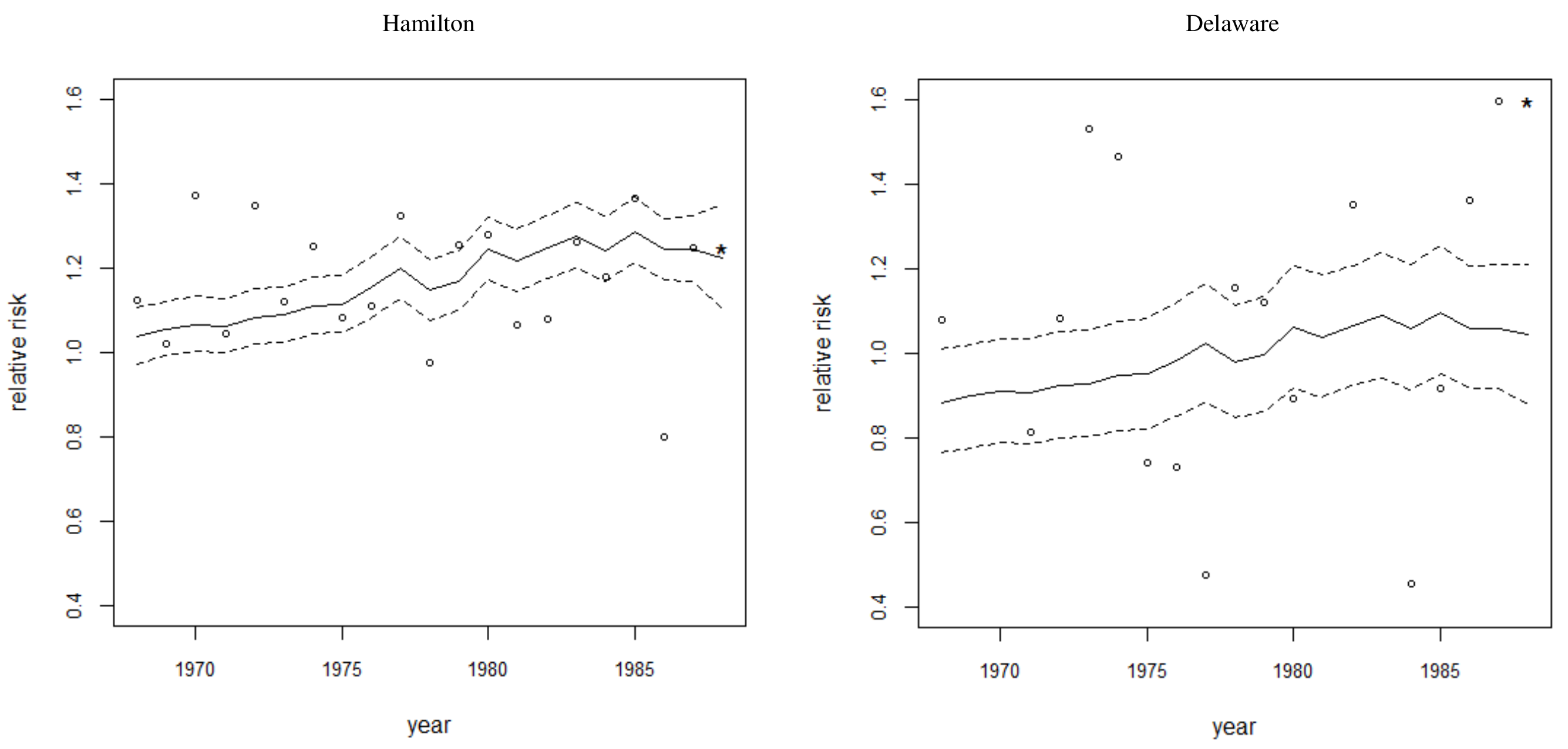}}
    \caption{The point estimates (solid lines) and 90\% interval estimates (dotted lines) of relative risks using $r_i^{(CG)}$ over time for two selected counties. In each figure, the circles denote the raw rates for year 1968 to 1977, and the ``$*$" denotes the raw rates for year 1988.}
    \label{f:trend}
\end{figure}

\section{Discussion}
\label{s:summary}

For specifying disease mapping models, we have argued that the usual internally standardized (IS) model is not generative, incoherent and tends to provide overestimation of uncertainty in inference. Toward this issue, we have proposed a coherent generative (CG) model, which models incidence, $p_i$, instead of relative risk, $r_i$, to avoid internal standardization. The generative model enables two relative risk estimators.  With regard to smoothing, estimators from the internally standardized model as well as the two estimators using the coherent specification yield essentially indistinguishable results.  We provide some analytical support in this regard as well as a simulation study. However, the posterior estimator for the relative risk, as a parametric function under the coherent model, tends to provide tighter credible intervals than that for the IS model both in simulation studies and the real data analysis. The CG model can be extended to richer model settings.  We illustrated with a simple dynamic version; its benefit in estimation of uncertainty still holds.

A future direction is to extend the single disease modeling to multiple diseases modeling. For example, let $p_{ij}$ denote the incidence rate for disease $j$ in region $i$. The logit transformation of $p_{ij}$ can be modeled as $\mbox{logit}(p_{ij})=\bX_{ij}^{\prime}\beta_{j}+\phi_{ij}$, where $\phi_{ij}$'s are spatial random effects with a multivariate conditional autoregressive (MCAR) prior.  Another future possibility is to employ the richer CAR-type models mentioned in the Introduction with the CG specification.


\backmatter






%
\bibliographystyle{biom}
\bibliography{reference}





\appendix
\section{}
\subsection{Posterior computation for the coherent generative model \eqref{eq:CG_model}}
Letting $\bm{\phi}=(\phi_1,\phi_2,...,\phi_I)^{T}$, the joint posterior is
\begin{equation}
f(\bm{\phi},\beta,\tau \mid \bm{n},\bm{Y},\bm{X},a,b)
\propto\{\overset{I}{\underset{i=1}{\prod}}f(Y_i\mid n_{i},\bX_i,\beta,\phi_{i})\}
f(\bm{\phi} \mid \tau)
f(\tau \mid a,b) f(\beta).
\end{equation}
The likelihood function is
\begin{equation}
\prod_{i=1}^{I}
\exp\left[-\frac{n_{i}}{1+\exp\{-(\bX_i^{\prime}\beta +\phi_{i})\}}\right]
\left[\frac{n_{i}}{1+\exp\{-(\bX_i^{\prime}\beta+\phi_{i})\}}\right]^{Y_i}/Y_i!.
\end{equation}
The resulting full conditional distributions are as follows:
\begin{enumerate}
  \item[a.] $f(\phi_{i} \mid \cdot)
    \propto \exp(-\frac{\tau w_{i+}}{2}(\phi_{i}-\bar{\phi_{i}})^2
    -\frac{n_{i}}{1+\exp\{-(\bX_i^{\prime}\beta+\phi_{i})\}}
    -Y_i\log[1+\exp\{-(\bX_i^{\prime}\beta+\phi_{i})\}])$, where $w_{i+}=\sum^{I}_{j=1}w_{ij}$ and $\bar{\phi_{i}}=\sum_j w_{ij} \phi_j / w_{i+}$;
  \item[b.] $f(\beta\mid \cdot)
    \propto \exp (-\sum_{i=1}^{I}\frac{n_{i}}{1+\exp\{-(\bX_i^{\prime}\beta +\phi_{i})\}}
    +\sum_{i=1}^{I}Y_i\log[\frac{n_{i}}{1+\exp\{-(\bX_i^{\prime}\beta +\phi_{i})\}}])$;
  \item[c.] $f(\tau \mid \cdot)\sim \mbox{Gamma}\{a+\frac{I}{2},
    b+\frac{1}{2} \sum^{I}_{i=1} \sum_{j<i}w_{ij}(\phi_i-\phi_j)^2 \}$.
\end{enumerate}

For $\phi_i, i=1,2,..,I$ and $\beta$, we use a Metropolis step to update. Specifically, we begin with a univariate version of this algorithm, in which the proposal distribution for each target variable is a univariate normal candidate density centered at the current value and having some variance $\sigma^2$. Then $\sigma^2$ is tuned to provide the Metropolis acceptance ratio between 15\% to 40\%. 

\end{document}